
\magnification=1200
\tolerance=100000
\overfullrule=0pt
\baselineskip=14pt
\def\lsim{<\kern-2.5ex\lower0.85ex\hbox{$\sim$}\ }
\def\rsim{>\kern-2.5ex\lower0.85ex\hbox{$\sim$}\ }

\centerline{\bf Self-adjoint Extension Approach to the spin-1/2
Aharonov-Bohm-Coulomb Problem}
\centerline{D. K. Park\footnote *{On leave from Department
of Physics, Kyung Nam University, Masan, 631-701, Korea}}
\centerline{Department of Physics and Astronomy}
\centerline{University of Rochester, Rochester, NY  14627, USA}
\bigskip
\noindent Abstract

The spin-1/2 Aharonov-Bohm problem is examined in the Galilean
limit for the case in which a Coulomb potential is included.  It
is found that the application of the self-adjoint extension method
to this system yields singular solutions only for one-half the
full range of flux parameter which is allowed in the limit of
vanishing Coulomb potential.  Thus one has a remarkable example
of a case in which the condition of normalizability is necessary
but not sufficient for the occurrence of singular solutions.
Expressions for the bound state energies are derived.  Also the
conditions for the occurrence of singular solutions are obtained
when the non-gauge potential is $\xi/r^p (0\leq p< 2)$.

\noindent I.  Introduction

In classical electrodynamics, the scalar and vector potentials are
merely a convenient tool for the calculation of fields.  In quantum
mechanics, however,
Aharonov and Bohm$^1$ (AB) gave a physical significance to the
vector potential, a result which has led to
 many theoretical and
experimental attempts$^2$ to establish the AB effect.

Recently the spin-1/2 AB problem has been of interest in various branches of
theoretical physics.  For example it appears in the interaction between
matter and cosmic strings.$^{3,4}$  It also appears in the anyonic
theory.$^5$

The magnetic flux tube, which shows up in
the AB problem has been treated in different ways..  Hagen$^{6,7}$ chose the
physically motivated expression
$$H \propto \lim_{R\to 0} {1\over R} \delta(r-R)\eqno(1.1)$$
where $H$ is the magnetic field and used it to
examine the validity of requiring
solutions to be regular at the origin as done by Aharonov and Bohm.$^1$
 He applied his method$^8$ to the spin-1/2 AB scattering
problem and concluded that the elimination of the singular solution
{\it ab initio} is not valid.  Using the expansion of the
upper component of the Dirac field
$$\psi_1(r,\theta) = \sum^\infty_{m = -\infty} f_m (r) e^{im\theta}
\eqno(1.2)$$
he calculated the solutions of the radial Schr\"odinger equation
in the $r<R$ and $r>R$ regions separately.  Upon applying the boundary
conditions at $r=R$ in the $R\to0$ limit, he derived
the fact that for the case
$$\mid m\mid + \mid m + \alpha \mid = - \alpha s\eqno(1.3)$$
$$\mid m + \alpha \mid < 1\; ,\eqno(1.4)$$
where
$$\eqalign{\alpha &= - e \int^\infty_0 dr\; rH(r)\cr
s &= \cases {1  &for spin up\cr
             -1 &for spin down\cr}\cr}\eqno(1.5)$$
only the solution singular at the origin
contributes to the radial wave function.  This contribution of the
singular solution gives a non-trivial scattering amplitude.

There was another more mathematical approach which was carried out
by Gerbert.$^9$  He chose the expression of the magnetic
flux tube to be
$$H \propto {1\over r}\delta(r)\eqno(1.6)$$
and imposed (1.4) as a normalizability condition.
He  then applied the self-adjoint extension$^{10,11}$ method to the
partial wave satisfying (1.4).
 Since the self-adjoint extension method gives
a one-parameter family of solutions, his scattering amplitude contains
a self-adjoint extension parameter.
His results coincide with those of Hagen when the self-adjoint
extension parameter, say $\theta$, equals $\pi/4$.  Thus for
$\theta = \pi/4$ only singular solutions contribute to the
radial wave function.  Subsequently Jackiw$^{12}$ gave some additional
insight into the self-adjoint extension formalism.  He proved that
the self-adjoint extension formalism gives a result identical to that
of the renormalization method when the
potential is a delta-function if a certain
relation between the self-adjoint extension parameter and the renormalized
coupling constant is satisfied.
Although Gerbert obtained the condition (1.4) by invoking normalizability
it will be shown that the condition of normalizability is only
necessary but not sufficient for the occurrence of singular solutions.
This statement can be proved by applying the self-adjoint extension
method to the spin-1/2 AB problem but including also a
 Coulomb potential $V(r) = \xi/r$.

Recently the spin-1/2 Aharonov-Bohm-Coulomb (ABC) problem was analyzed
by Hagen.$^{13}$  He showed that this system yields
singular solutions only for one-half the full range which is allowed in
the limit of vanishing Coulomb potential, namely,
$$\mid m + \alpha \mid < {1\over 2}\eqno(1.7)$$
and also obtained the bound state energies
$$\varepsilon_n = - {1\over 2} {M\xi^2\over [n - {1\over 2} \pm
\mid m + \alpha \mid ]^2} \qquad n = 1,2,\ldots\eqno(1.8)$$
where the upper and lower signs refer respectively to the case of
regular and irregular solutions.

In the present paper  the ABC problem is
 analyzed in the context of the the self-adjoint
extension method.
  In order to avoid complications associated
with Klein's paradox$^{14}$ the analysis is
performed within the framework of the Galilean$^{15}$ spin-1/2 wave
equation.  It will be shown that the ABC system is a striking example of a
case in which the condition of normalizability is a necessary
but not a sufficient condition
for the occurrence of singular solutions.
It will also be proved that the condition for the occurrence of singular
solutions does not coincide with that of normalizability in the more general
case $V(r) = \xi/r^p (0<p<2)$.  The
sufficiency conditions are derived when this general potential is included
in the AB system.
Before proceeding it is important to state somewhat more explicitly
the reasoning which motivates the use of the Galilean limit.
Insofar as numerical calculations one concerned one clearly expects
that limit to be useful only in a domain in which the binding energy
of a state is much less than the rest energy. For scattering states
a realistic theory must be capable of dealing with particle
creation processes as well. Thus the reliability of any wave
equation(whether Galilean or special relativistic) is again limited
in energy. The essential problem is that the unbounded nature of
the $1/r$ potential gives rise to difficulties related to the
Klein's paradox even when calculating at relatively low energies.
The Galilean equation has no such defect and is thus an ideal tool
for the study of the ABC problem.

This paper is organized as follows.  In Section II the ABC problem is
solved in the $r\neq 0$ region.  The necessary conditions for
the occurrence of
singular solution are obtained.  In Sec. III the self-adjoint extension
method is applied to the restricted subspace obtained in Sec. II.  It is
shown that the self-adjoint extension of the Hamiltonian gives the boundary
conditions at the origin and that
these boundary conditions reduce the possible range of $\alpha$
for the occurrence of singular solutions.  Also the expressions for the
bound state energies are derived.  In Section IV the condition for
solutions singular at the origin is examined when the potential
$\xi/r^p (0 \leq p < 2)$ is included in the AB problem.  In Sec. V a
brief conclusion is given.
\bigskip
\noindent II. Solution of the ABC Problem in the $r\neq 0$ Region

In this section the ABC problem is solved in the $r\neq 0$ region within
the framework of the Galilean limit.  We start with the Dirac equation
$$[m\beta + \beta\gamma_i\Pi_i]\psi = E\psi\eqno(2.1)$$
where
$$\eqalign{\Pi_i &= - i\partial_i - eA_i\; ,\cr
\beta &= \sigma_3\; , \cr
\beta\gamma_i &= (\sigma_1, s \sigma_2)\;,\cr}\eqno(2.2)$$
and $s$ is given in Eq. (1.5).  If one takes the Galilean limit
$$\eqalign{E &= M + {\cal E}\cr
M & \gg {\cal E}\cr}\eqno(2.3)$$
and includes the Coulomb potential by letting ${\cal E} \to
{\cal E} - {\xi\over r}$, Eq. (2.1) becomes
$$ \left\lgroup\matrix {{\cal E}-{\xi\over r},
&-(\Pi_1 - is\Pi_2)\cr -(\Pi_1 + is\Pi_2),&2M \cr}\right\rgroup
\left\lgroup\matrix{\psi_1\cr \psi_2\cr}\right\rgroup = 0\; .\eqno(2.4)$$
It may be worth noting that there is no contradiction between (2.3)
and the use of an unbounded Coulomb potential.  In fact once the free
Galilean wave equation has been derived, any potential consistent
with general Galilean invariance can be considered.
The magnetic flux tube is specified by
$$\eqalign{eA_i &= \alpha \epsilon_{ij} {r_j\over r^2}\; , \cr
eH &= - \alpha \delta(r)/r \; ,\cr}\eqno (2.5)$$
where $\epsilon_{ij} = - \epsilon_{ij}$ and $\epsilon_{12} = + 1$.
By using Eq. (2.5) the Schr\"odinger equation of $\psi_1$ is
easily derived
$$H \psi_1 = {\cal E} \psi_1\eqno(2.6)$$
where
$$\eqalign{H &= H_0 + {\alpha s\over 2Mr} \delta (r)\; ,\; ,\cr
& {\rm and} \cr
H_0 &= {1\over 2M} \left[\left({1\over i} \vec\nabla - e \vec A\right)^2
+ {2M\xi\over r}\right]\; .\cr}\eqno(2.7)$$
If one decomposes the fermion field as
$$\psi (r,\theta) = \sum^\infty_{m = -\infty}
\pmatrix{\chi_{1,m}(r)\cr \chi_{2,m}(r)\cr}e^{im\theta}\; , \eqno(2.8)$$
the Schr\"odinger equation for $\chi_{1,m}(r)$ becomes
$$\left[{d^2\over dr^2} + {1\over r} {d\over dr} + k^2 -
{(m+\alpha)^2\over r^2} - {2M\xi\over r} - \alpha s \delta (r)/r
\right] \chi_{1,m}(r) = 0 \eqno(2.9)$$
where $k^2 = 2M {\cal E}$, and $\chi_{2,m}(r)$ is derived from
$\chi_{1,m}(r)$ by
$$\chi_{2,m}(r) = - {i\over 2M}\left({d\over dr} -
{(m+\alpha)s\over r}\right) \chi_{1,m}(r)\; .\eqno(2.10)$$
By directly solving Eq. (2.9) the solutions for $\chi_{1,m}(r)$ in the
$r\neq 0$ region is derived
$$\eqalign{\chi_{1,m}(r) &= A_me^{ikr}(-2ikr)^{m+\alpha}
F\left(m+\alpha + {1\over 2} + {iM\xi\over k} \mid 2(m+\alpha)+1\mid
-2ikr\right)\cr
&+B_m e^{ikr}(-2ikr)^{-(m+\alpha)}F \left( -(m+\alpha) +
{1\over 2} + {iM\xi\over k}\mid 1 - 2(m+\alpha)\mid -2ikr\right)\cr}
\eqno(2.11)$$
where $F (a\mid c\mid z)$ is usual confluent hypergeometric function.
By inserting Eq. (2.11) into Eq. (2.10) one obtains
$$\eqalign{ \chi_{2,m}(r) &=\cr
& -A_me^{ikr}\Bigl[ {(m+\alpha + {1\over 2})^2 + ({M\xi\over k})^2
\over 4M(m+\alpha +1)
     (m+\alpha+{1\over 2})^2} (-2ikr)^{m+\alpha+1}
     F\bigl( m+\alpha + {3\over 2} + {iM\xi\over k}\big\vert\cr
&\quad 2(m +\alpha)+3\mid-2 ikr\bigr)
              + {i\xi\over 2(m+\alpha)+1}(-2ikr)^{m+\alpha}\cr
&\quad\times F\bigl(m+\alpha + {1\over 2} + {iM\xi\over k}\big\vert 2(m+\alpha)
           +1\mid -2 ikr\bigr)\Bigr]\cr
& +B_me^{ikr}\Bigl[ {2k(m+\alpha)\over M} (-2ikr)^{-(m+\alpha)-1}
    F(-(m+\alpha)-{1\over 2} + {iM\xi\over k}\big\vert\cr
&\quad -2(m+\alpha)-1\big\vert-2ikr) - {i\xi\over 2(m+\alpha)+1}
    (-2ikr)^{-(m+\alpha)}\cr
&\quad \times F(-(m+\alpha) + {1\over 2} + {iM\xi\over k}\big\vert
    1-2(m+\alpha)\big\vert -2ikr)\Bigr]\cr
&\qquad\qquad\qquad\qquad\qquad\qquad\qquad\qquad\qquad ({\rm for}\; s=1)\cr
& A_me^{ikr}\Bigl[ -{2k(m+\alpha)\over M}(-2ikr)^{m+\alpha -1}
    F(m+\alpha - {1\over 2} + {iM\xi\over k} \big\vert\cr
&\quad 2(m+\alpha)-1\big\vert -2ikr) + {i\xi\over 2(m+\alpha)-1}
   (-2ikr)^{m+\alpha}\cr
&\quad\times F(m+\alpha + {1\over 2} + {iM\xi\over k}\big\vert
2(m+\alpha ) + 1 \big\vert -2ikr)\Bigr]\cr
& + B_me^{ikr} \Bigl[ {(m+\alpha - {1\over 2})^2 + ({M\xi\over k})^2\over
4M(m+\alpha -1)(m+\alpha - {1\over 2})^2}(-2ikr)^{-(m+\alpha)+1}\cr
&\qquad \times F(-(m+\alpha) + {3\over 2} + {iM\xi\over k}\big\vert\cr
&\quad -2(m+\alpha)+3\big\vert -2ikr) + {i\xi\over 2(m+\alpha)-1}
    (-2ikr)^{-(m+\alpha)}\cr
&\quad \times F(-(m+\alpha)+{1\over 2} + {iM\xi\over k}\big\vert
    -2(m+\alpha)+1\big\vert -2ikr)\Bigr]\cr
&\qquad\qquad\qquad\qquad\qquad\qquad\qquad\qquad\qquad ({\rm for}\; s=-1)
\; . \cr} \eqno(2.12)$$
Either $A_m$ or $B_m$ must be zero by the condition of normalizability
except in the subspace
$$\eqalign{s &= 1 \qquad\quad m = -N-1\;,\cr
s &= -1 \qquad m = -N\;,\cr}\eqno(2.13)$$
where
$$\eqalign{\alpha &= N + \beta\; ,\cr
N &; {\rm integer}\;,\cr
0 &< \beta < 1\;.\cr}\eqno(2.14)$$
It is easy to see that
 Eq. (2.13) is merely a different description of Eqs. (1.3) and (1.4). From
 Eq. (2.11) and (2.12) it follows that
$\chi_{1,m}(r)$ and $\chi_{2,m}(r)$ cannot both be chosen to be regular
solutions when the condition (2.13) is satisfied.  The next section
will analyze the ABC problem by the self-adjoint extension method and
obtain the result
 that the condition (2.13) is necessary but not sufficient for
the occurrence of singular solution.
\bigskip

\noindent III. Self-adjoint Extension

In Sec. II it was shown that $\chi_{1,m}(r)$ and $\chi_{2,m}(r)$
cannot both be chosen as regular solutions when the condition (2.13) is
satisfied.  This means that the Hamiltonian of $\chi_{1,m}(r)$
$$\eqalign{H &= H_0 + {\alpha s\over 2Mr} \; \delta(r)\cr
\noalign{\vskip4pt}
H_0 &= - {1\over 2M} \left[ {d^2\over dr^2} + {1\over r} {d\over dr}
- {(m+\alpha)^2\over r^2} - {2M\xi\over r}\right]\cr}\eqno(3.1)$$
is not a self-adjoint operator.  In order for the Hamiltonian to be a
self-adjoint operator the domain of definition of $H_0$ has to be
extended to the deficiency subspace of $H_0\mid_{m=-N-1\; {\rm or}
\; m=-N}$ which is spanned by the solutions of the differential
equations,
$$\left[{d^2\over dr^2} + {1\over r}{d\over dr} - {(m+\alpha)^2\over r^2}
\pm ik^2 - {2M\xi\over r}\right] \chi_\mp(r) = 0\; .\eqno(3.2)$$
This means that the boundary condition at the origin
$$\lim_{r\to 0} r^{\mid m +\alpha\mid} \chi (r) = \lambda \lim_{r\to 0}
{1\over r^{\mid m + \alpha \mid}} \left[ \chi (r) -
\left ( \lim_{r^\prime \to 0} r^{\prime\; \mid m + \alpha \mid}
\chi (r^\prime)\right ) {1\over r^{\mid m + \alpha \mid}} \right]\; ,
\eqno(3.3)$$
where $\lambda$ is the self-adjoint extension parameter, must be required for
 $\chi_{1,m}(r)$.  To treat both cases of Eq. (2.13)
simultaneously an expression for $\chi_{1,m}(r)$ which is
different from Eq.(2.11)
is more convenient, namely,
$$\eqalign{\chi_{1,m}(r) &= A_me^{ikr} (-2ikr)^{\mid m + \alpha\mid}
F(\mid m + \alpha \mid + {1\over 2} + {iM\xi\over k} \big\vert
 2 \mid m + \alpha\mid + 1\big\vert - 2ikr)\cr
&\quad + B_m e^{ikr} (-2ikr)^{-\mid m + \alpha \mid}
  F ( -\mid m + \alpha \mid + {1\over 2} + {iM\xi\over k} \big\vert 1 - 2
  \mid m + \alpha \mid \big\vert -2 ikr)\; . \cr}\eqno(3.4)$$
Note that $A_m$ and $B_m$ are the coefficients of the regular and
irregular solutions respectively.  By inserting Eq. (3.4) into the boundary
condition (3.3) the following relation between the coefficients $A_m$ and
$B_m$ is derived.
$$\eqalign{\lambda (-2ik)^{\mid m + \alpha \mid} A_m &=
 (-2ik)^{-\mid m + \alpha \mid} B_m \Bigl[ 1 - \cr
&\quad   {2 \lambda M \xi\over 1 - 2\mid m + \alpha \mid}
\bigl(\lim_{r\to 0} r ^{1 - 2\mid m + \alpha \mid}\bigr)
\Bigr]\cr}\eqno(3.5)$$
Note that the coefficient of $B_m$ diverges as
$\lim_{r\to 0} r^{1-2\mid m + \alpha \mid}$, if $\mid m + \alpha \mid
 > {1\over 2}$.
Thus $B_m$ is zero if $\mid m + \alpha \mid > {1\over 2}$, and the
condition for the occurrence of a singular solution is$^{16}$
$$\mid m + \alpha \mid < {1\over 2}\; .\eqno (3.6)$$
Since this is a subspace of Eq. (1.3) and (1.4),
 or equivalently Eq. (2.13), it is
seen that the condition of normalizability is only a necessary condition
for the occurrence of singular solution.  So let us restrict ourselves
to the subspace (3.6).

Since the bound states are obtained in the positive imaginary region
of $k$, one can derive the bound state from Eq. (3.4) by changing
$k \to i\sqrt {2MB}$, where $B = - \varepsilon$ is the bound state energy;

$$\eqalign{\chi^B_{1,m}(r) &= A_me^{-\sqrt{2MB} r}
   (2\sqrt {2MB} r)^{|m+\alpha|}\cr
&\quad \times F \bigl(|m+\alpha| + {1\over 2} +
   \sqrt{M\over 2B}\xi \big\vert 2|m+\alpha| + 1 \big\vert
   2\sqrt{2MB}r\bigr)\cr
&\; + B_m e^{-\sqrt{2MB} r} (2\sqrt{2MB} r)^{-|m+\alpha |}\cr
&\quad \times F\bigl( -|m+\alpha| + {1\over 2} + \sqrt{M\over 2B}\xi \big\vert
1 - 2|m+\alpha | \big\vert 2\sqrt{2MB} r \bigr) \cr}\eqno(3.7)$$
However, Eq. (3.7) is still not
guaranteed to be a bound state.  In order to be a bound
state, $\chi^B_{i,m}(r)$ must be normalizable at large $r$. This
condition gives the relation
$$A_m {\Gamma (2\mid m + \alpha \mid + 1)\over
\Gamma (\mid m + \alpha \mid + {1\over 2} + \sqrt{M\over 2B}\xi)}
+ B_m {\Gamma (1 - 2\mid m + \alpha \mid )\over
\Gamma ({1\over 2} - \mid m + \alpha \mid + \sqrt{M\over 2B}\xi)}
= 0 \eqno (3.8)$$
By inserting Eq. (3.8) into Eq. (3.7) the bound state is then obtained
$$\chi^B_{1,m}(r) = {\cal N}_m {1\over \sqrt r} W_{-\sqrt{M\over 2B}
\xi , \mid m + \alpha\mid} (2\sqrt {2MB} r)\eqno (3.9)$$
where ${\cal N}_m$ is a normalization constant and $W_{a,b}(z)$ is
the usual Whittaker function.

Another relation between $A_m$ and $B_m$ is derived by inserting
Eq. (3.7) into the boundary condition (3.3)
$$\lambda (2\sqrt{2MB})^{\mid m + \alpha \mid} A_m -
(2\sqrt{2MB})^{-\mid m + \alpha \mid} B_m = 0. \eqno (3.10)$$
Thus the bound state energy is implicitly determined from
Eqs. (3.8) and (3.10) by the secular equation
$$\eqalign{(2\sqrt{2MB} &)^{-|m+\alpha|}
  {\Gamma (2|m+\alpha| + 1)\over \Gamma ({1\over 2} + |m+\alpha | +
  \sqrt{M\over 2B} \xi)}\cr
\noalign{\vskip4pt}
& + \lambda (2\sqrt{2MB})^{|m+\alpha|}
   {\Gamma (1 - 2|m+\alpha |)\over \Gamma ({1\over 2} - |m+\alpha | +
   \sqrt{M\over 2B}\xi )} = 0.\cr}\eqno(3.11)$$
Although Eq. (3.11) is too complicated to evaluate the bound state energies,
its limiting feature is interesting.  First in the $\lambda \to 0$ or
$\infty$ limit bound state energies are explicitly determined as the
poles of the gamma function
$$\eqalign{ \lim_{\lambda\to 0} B &= {M\over 2}
  {\xi^2\over [n - {1\over 2} + |m+\alpha |]^2}\; ,\cr
\noalign{\vskip4pt}
& \qquad\qquad\qquad\qquad\qquad\qquad n = 1, 2, \ldots \cr
\noalign{\vskip4pt}
\lim_{\lambda\to \infty} B &= {M\over 2}
  {\xi^2\over [n - {1\over 2} - |m + \alpha |]^2}\; .\cr}\eqno(3.12)$$
These coincide (as expected)
 with  Eq. (1.7) of Hagen.$^{13}$
Another interesting limit is the case of vanishing Coulomb potential.
In the $\xi \to 0$ limit a bound state energy is explicitly determined
from Eq. (3.11)
$$\lim_{\xi\to 0} B = {2\over M} \left[ - {\Gamma (1 + |m+\alpha |)
\over \lambda \Gamma (1 - |m + \alpha |)}\right]^{1\over |m+\alpha |}
\eqno(3.13)$$
which is the Galilean limit of Gerbert's$^9$ bound state energy.
In fact Eq. (3.13) coincides with Gerbert's result when
$$\lambda = - 2 \sqrt{E-M\over E+M} |m+\alpha |
{M^{1-2|m+\alpha |}\over k}
\tan \left({\pi\over 4} + {\theta\over 2}\right)\eqno(3.14)$$
and the relativistic relation $k^2 = E^2 - M^2$ is used.
\bigskip

\noindent IV.  Conditions for Singular Solutions when $\xi/r^p$ Potential
is Included

In this section yet another example is given which illustrates that the
conditions for singular solutions does not coincide with those of
normalizability.  Instead of the Coulomb potential one includes a
more general potential $\xi/r^p$ in Eq. (2.9).

Having shown already that self-adjoint extension method give the same
conditions for singular solutions as the method used in Ref. [13],
one uses here the latter approach.

To this end Eq. (2.9) with $\xi/r^p$ potential is divided into
$$ \left[ {d^2\over dr^2} + {1\over r} {d\over dr} -
{m^2\over r^2} + k^2_0\right] \chi_{1,m} (r) = 0 \qquad\qquad\quad
(r < R)\eqno(4.1a)$$
$$\left[ {d^2\over dr^2} + {1\over r} {d\over dr} + k^2 -
{(m+\alpha)^2\over r^2} - {2M\xi\over r^p}\right]\chi_{1,m} (r)=0\qquad
(r > R)\eqno(4.1b)$$
where
$$\eqalign{ k^2_0 &= 2M(\varepsilon - U_R)\cr
k^2 &= 2M\varepsilon\cr}\eqno(4.2)$$
and $U_R$ is the constant potential $\xi/R^p$.

One easily obtains $\chi_{1,m}(r)$ for $r<R$ region since it must be
regular at the origin. Thus
$$\chi_{1,m}(r) = C_mJ_{|m|}(k_0r)\; .\eqno(4.3)$$
Although the analytic solution of Eq. (4.1b) cannot be found easily, we
can evaluate the solution by an expansion about the
origin.  The first few terms are
$$\chi_{1,m}(r) = \cases {A_me^{ikr} r^{|m +\alpha |} (1-ikr + \ldots )\cr
\noalign{\vskip4pt}
+B_me^{ikr} r^{-|m+\alpha |} (1-ikr + {2M\xi\over (2-p)(2-p-2|m+\alpha|)}
r^{2-p} + \ldots)\cr
\noalign{\vskip4pt}
\qquad\qquad\qquad\qquad\qquad\qquad (0 \leq p < 1)\cr
\noalign{\vskip4pt}
A_me^{ikr} r^{|m+\alpha |}(1 + {2M\xi\over (2-p)(2-p+2|m+\alpha |)}
r^{2-p} + \ldots )\cr
\noalign{\vskip4pt}
+ B_me^{ikr} r^{-|m+\alpha |} (1 + {2M\xi\over (2-p)(2-p-2|m+\alpha |)}
r^{2-p} + \ldots )\cr
\noalign{\vskip4pt}
\qquad\qquad\qquad\qquad\qquad\qquad (1 < p < 2)\cr}\eqno(4.4)$$
where $A_m$ and $B_m$ are the coefficients of the regular and irregular
solutions respectively.

By the usual matching conditions
$$\eqalign{\lim_{\varepsilon \to 0^+} (\chi_{1,m}(R+\varepsilon) -
\chi_{1,m}(R-\varepsilon)) &= 0\; ,\cr
\lim_{\varepsilon \to 0^+}{d\over dr} (\chi_{1,m}(R+\varepsilon) -
\chi_{1,m}(R-\varepsilon)) &= {\alpha s\over R} \chi_{1,m}(R)\;,
\cr}\eqno(4.5)$$
one gets the relation between $A_m$ and $B_m$.  In the limit of
small $R$ it becomes
$${A_m\over B_m} = - {|m|+|m+\alpha | + \alpha s\over |m|-|m+\alpha |
+\alpha s} R^{-2|m+\alpha |} + {\cal O}(R^{-2|m+\alpha |+2-p})\eqno(4.6)$$
for $0 \leq p < 2$.

{}From (4.6) one gets the conditions for the singular solutions
$$\eqalign{|m|+|m+\alpha | &= - \alpha s\cr
|m+\alpha | &> 1 - {p\over 2}\;. \cr}\eqno(4.7)$$
Thus, except for the trivial case $p=0$ the
conditions for singular solutions do not coincide
with the normalizability
condition $|m+\alpha | < 1$.
\bigskip

\noindent V.  Conclusion

The self-adjoint extension method has been applied
here to the spin-1/2 ABC problem.
It was shown that the self-adjoint extension method yields
$|m+\alpha | < {1\over 2}$
as the condition for the occurrence of singular solution.
This is in agreement with the results of Ref. [13] which were
obtained by a different method.  It has also been shown
that the condition of normalizability is necessary but not sufficient
for the occurrence of singular solutions.  Expressions for the bound
state wave function and bound state energy have been derived as well.
For the $\lambda \to 0$ and
 $\lambda \to\infty$ limits
these bound states have been seen to coincide with those of the regular and
irregular cases respectively of Ref. [13].  Also the $\xi \to 0$ limit of the
Galilean version of the bound state energy of Ref. [9] was obtained.
Finally it was found that the condition for the existence of singular
solutions is $\mid m + \alpha \mid > 1 - p/2$ when the general
non-gauge potential $\xi/r^p$ is included in the AB problem. This result
implies that for a gas of such particles the discontinuities in the
second virial coefficient$^{17}$ are shifted from integer values of
$\alpha$ for the Coulomb potential to $n \pm {p\over 2}$ values ($n$ is
an integer).  As in the Coulomb case the transisition point has no
dependence on the strength of the potential.

\bigskip

\noindent Acknowledgements

I am very grateful to C. R. Hagen for introducing the subject and for
various crucial suggestions.  This work is partially supported by the
Korean Science and Engineering Foundation.
\medskip

\noindent References
\item{[1.]} Y. Aharonov and D. Bohm, Phys. Rev. {\bf 115}, 485 (1959).
\item{[2.]} M. Peshkin and A. Tonomura, {\it The Aharonov-Bohm Effect},
(Springer-Verlag, Berlin, 1989).
\item{[3.]} M. G. Alford and F. Wilczek, Phys. Rev. Lett. {\bf 62},
1071 (1989).
\item{[4.]} M. G. Alford, J. March-Russell and F. Wilczek, Nucl. Phys.
{\bf B328}, 140 (1989).
\item{[5.]} Y. H. Chen, F. Wilczek, E. Witten and B. I. Halperin, Int.
J. Mod. Phys. {\bf B3}, 1001, (1989).
\item{[6.]} C. R. Hagen, Phys. Rev. {\bf D41}, 2015 (1990).
\item{[7.]} C. R. Hagen, Int. J. Mod. Phys. {\bf A6}, 3119 (1991).
\item{[8.]} C. R. Hagen, Phys. Rev. Lett. {\bf 64}, 503, (1990).
\item{[9.]} Ph. de Sousa Gerbert, Phys. Rev. {\bf D40}, 1346 (1989).
\item{[10.]} A. Z. Capri, {\it Nonrelativistic Quantum Mechanics},
(Benjamin/Cummings, California, 1985).
\item{11.]} S. Albeverio, F.Gesztesy, R. Hoegh-Krohn and H. Holden,
{\it Solvable Models in Quantum Mechanics}, (Springer-Verlag, Berlin
1988).
\item{[12.]} R. Jackiw, ``Delta function potential in two- and
three-dimensional quantum mechanics", in M. A. B. Beg memorial
volume, A. Ali and P. Hoodbhoy, eds., (World Scientific,
Singapore, 1991).
\item{[13.]} C. R. Hagen, Phys. Rev. {\bf D48}, 5935 (1993).
\item{[14.]} J. D. Bjorken and S. D. Drell, {\it Relativistic
Quantum Mechanics}, (McGraw-Hill, New York, 1964).
\item{[15.]} J. M. L\'evy-Leblond, Commun. Math. Phys. {\bf 6}, 286 (1967).
\item{[16.]} Note that it is possible in principle to obtain a solution
for the domain $1/2 < \mid m +\alpha \mid < 1$ if (3.3) is replaced
by
$$\lim_{r\to 0} r^{\mid m +\alpha\mid} \chi (r) = \lambda \lim_{r\to 0}
{1\over r^{1 - \mid m +\alpha \mid}} \left[ \chi(r) -
\left ( \lim_{r^\prime \to 0} r^{\prime\; \mid m + \alpha \mid}
\chi (r^\prime)\right ) {1\over r^{\mid m + \alpha \mid}} \right]$$
and if one simultaneously performs a fine tuning on the extension
parameter $\lambda$. Since the latter is contrary to the spirit of
the self-adjoint extension method(which places no conditions on
the extension parameter(s)) this approach is not considered in
the present work.
\item{[17.]} T. Blum, C. R. Hagen, and S. Ramaswamy, Phys. Rev. Lett.
{\bf 64}, 709 (1990).

\vfil\eject
\end